\documentclass[fleqn,10pt]{wlscirep}
\usepackage{mathtools}
\usepackage[final]{pdfpages} % Include PDF

\title{Sustainability of Transient Kinetic Regimes and Origins of Death}

\author[1,*]{Dmitry Yu. Zubarev}
\author[2]{Leonardo A. Pach\'on}
\affil[1]{Department of Chemistry and Chemical Biology, Harvard University, Cambridge, MA 02138 USA.}
\affil[2]{Grupo de F\'isica At\'omica y Molecular, Instituto de F\'{\i}sica,
Facultad de Ciencias Exactas y Naturales,
Universidad de Antioquia UdeA; Calle 70 No. 52-21, Medell\'in, Colombia.}

\affil[*]{zubarev@fas.harvard.edu}

\begin{abstract}
It is generally recognized that a distinguishing feature of life is its peculiar capability to avoid equilibration. 
The origin of this capability and its evolution along the timeline of abiogenesis is not yet understood. 
We propose to study an analog of this  phenomenon that could emerge in non-biological systems. 
To this end, we introduce the concept of sustainability of transient kinetic regimes. 
This concept is illustrated via investigation of cooperative effects in an extended system of compartmentalized 
chemical oscillators under batch and semi-batch conditions. 
The computational study of a model system shows robust enhancement of lifetimes of the decaying 
oscillations which translates into the evolution of the survival function of the transient non-equilibrium regime. 
This model does not rely on any form of replication. 
Rather, it explores the role of a structured effective environment as a contributor to the system-bath 
interactions that define non-equilibrium regimes. We implicate the noise produced by the effective 
environment of a compartmentalized oscillator as the cause of the lifetime extension.
\end{abstract}

\begin{document}

\flushbottom
\maketitle
\thispagestyle{empty}

%\noindent Please note: Abbreviations should be introduced at the first mention in the main text ? no abbreviations lists. Suggested structure of main text (not enforced) is provided below.

\section*{Introduction}

Between ill-conditioned and ill-posed problems, the origin of life must be in a class of its own. 
Life on Earth is the only type of life known to us, its chemical basis is very constrained \cite{Ref01}, 
and information about the initial condition of prebiological evolution has been lost. 
Attempts to solve this problem have evolved along two complementary routes. 
One includes experimental and observational efforts to find connections between prebiological 
chemistry and biochemistry. 
It yielded multiple insights ranging from Urey-Miller experiment \cite{Ref02} to the recent cyano-sulfidic 
scenario \cite{Ref03} and prompted active search for habitable exoplanets \cite{Ref04}. 
The other is dedicated to establishing driving forces of abiogenesis and abstraction of the general 
evolutionary principles from their known biochemical realizations. 
Examples of such efforts include, but are not limited to, development of the theory of hypercycles 
and quasispecies \cite{Ref05} and evolution of chemical kinetics into population dynamics \cite{Ref06}, thermodynamic 
foundations of cellular metabolism \cite{Ref07} and replication \cite{Ref08}, information transfer in mutually catalytic 
systems \cite{Ref09} and, of course, the general area of non-equilibrium thermodynamics \cite{Ref10,Ref11,Ref12,Ref13}. 

A hallmark of life that strongly pushes for the formulation of a general evolutionary principle is its 
unique position with respect to the equilibrium. 
One way to express it is to borrow a quote from Schr{\"o}dinger \cite{Ref14}: ``It is by avoiding the rapid decay into the inert 
state of 'equilibrium' that an organism appears so enigmatic''. The enigma of ``avoiding the rapid 
decay'' is intimately related to the enigma of death as the event that has to be avoided. 
Unlike the origin of life, cellular death is ubiquitous, can be replicated and systematically studied. 
Aside from the catastrophic events, it involves elaborate regulatory processes that recruit the genetic, 
metabolic and enzymatic systems in what is called ``programmed cell death'' \cite{Ref15,Ref16}. 
Moving back in time to the origin of the respective biochemical mechanisms, how much would the 
notion of death change? What are the prebiological precursors of the evolved mechanisms of cell 
death? 

Given generality of the question, the ambiguity of the answer is unavoidable. 
Our general motivation is to identify phenomena that look similar to cell death in prebiological 
context in the absence of the evolved biochemical systems. 
If identified, they can serve as a foundation for analogical models \cite{Ref17,Ref18} of chemical evolution 
and enable studies of life-like physical-chemical systems that are not necessarily prebiologically 
plausible. 
We think of life in terms of non-equilibrium chemical processes with a peculiar spatio-temporal 
organization that can be maintained over a finite time. 
From this point of view a living organism exists in a transient regime -  a regime encountered 
on the system's way to the state that can be maintained for as long as the respective external 
conditions are maintained, such as thermodynamic equilibrium, chemical equilibrium, or a steady state. 
What makes life unique in the sense of the above-mentioned quote, is that there exist mechanisms 
that extend the lifetime of this transient regime well beyond what one would expect from a typical 
non-equilibrium system decaying exponentially fast. 
Such mechanisms are either intrinsic, i.e., contained within the system, and their engagement is 
inevitable, or they come from some special structure of the environment. 
The latter is possible but difficult to address. 
Our conjecture is that death comes into play in prebiological world as a point of failure of the intrinsic 
mechanisms that ensure sustainability of the transient kinetic regimes. 

\begin{figure*}[htp!]
            \includegraphics[width=0.45\textwidth]{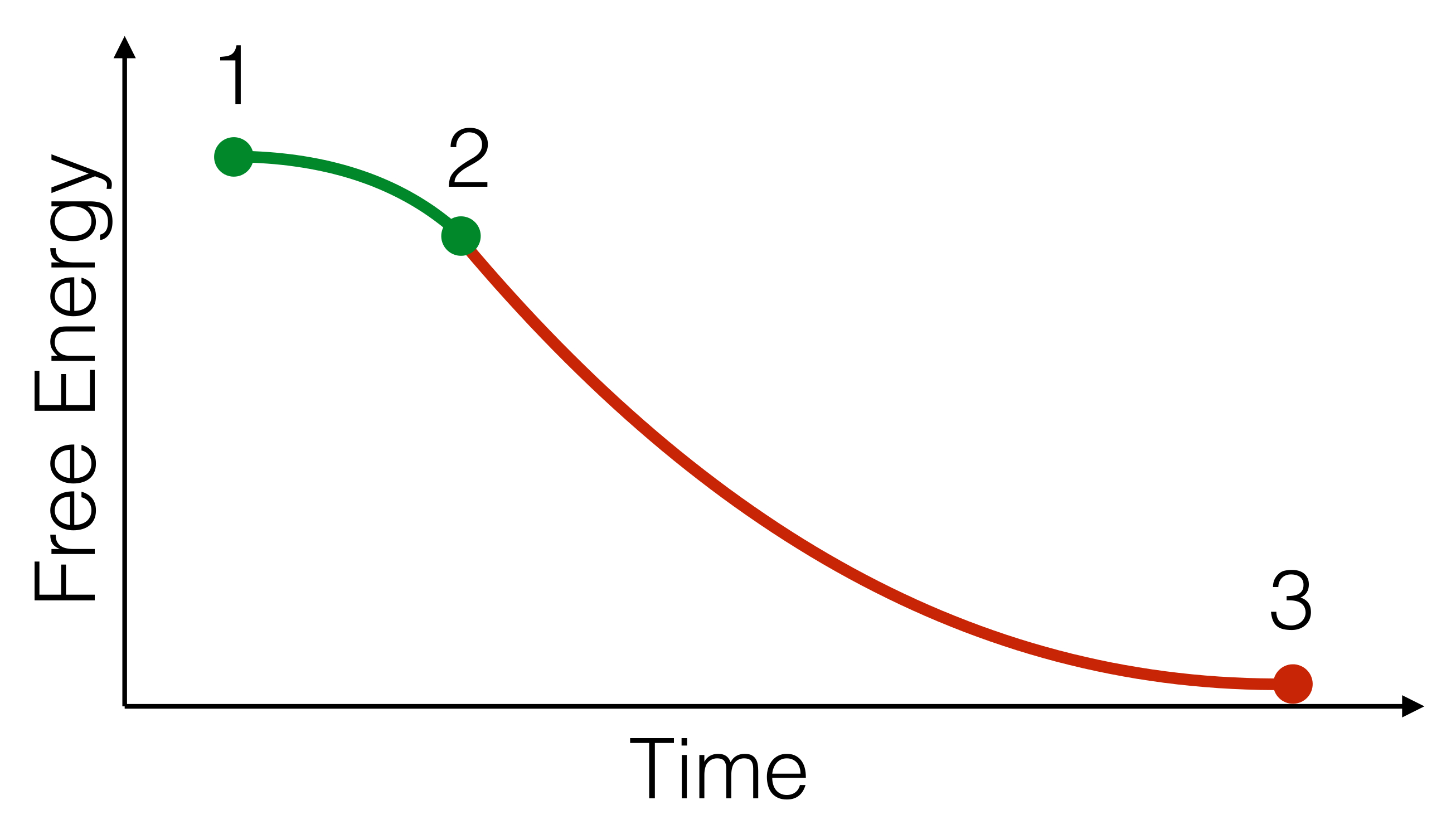} 
\caption{Schematic representation of the time-evolution of a living system. 
Point 1 is chosen arbitrarily with respect to the birth event.  It belongs to an identifiable dynamic regime of time-evolution
associated with non-equilibrium conditions. We recognize such a regime as life. 
Point 2 labels the event of death. It is identified as the loss of the aforementioned regime and occurs away from the equilibrium. Point 3 labels the equilibrium that can be chemical or thermodynamic depending on the specifics of the problem.  
The entire segment 1-3 corresponds to the equilibration process, including the segment 1-2. 
We refer to the segment 1-3 as a ``transient process''; 
as a stage of the transient process, we refer to the segment 1-2 as a ``transient regime''. The lifetime of the transient regime is finite.}
   \label{fig:SchmtcRpsnttn}
\end{figure*}

Figure 1 shows a schematic representation of the stages of time-evolution of a living system. Here, we refer to the entire time-evolution from some non-equilibrium state (Point 1, Fig.~\ref{fig:SchmtcRpsnttn}) to the equilibrium (Point 3, Fig.~\ref{fig:SchmtcRpsnttn}) as a \textit{transient process} (segment 1-3 in Fig.~\ref{fig:SchmtcRpsnttn}). 
In this study, we will think about life as a \textit{transient regime} (segment 1-2 in Fig.~\ref{fig:SchmtcRpsnttn}) which is a part of the transient process. As a stage of the equlibration process, the transient regime has some unique features that make it identifiable. 
The loss of the transient regime in Point 2 corresponds to death. 
At this point the system is still away from the equilibrium -- for example, hydrolysis of components of living cells, such as nucleic acids, will proceed on the time scale of years \cite{Ref19} outside living cells. The equilibration process continues until the equilibrium is reached (segment 2-3 in Fig.~\ref{fig:SchmtcRpsnttn}). 
As far as avoiding rapid equilibration, the enigma of life comes from the mechanisms 
that extend the lifetime of the transient regime 1-2 and delay transition from the segment 1-2 
to the segment 2-3.
Sustainability of the transient regime, therefore, is the capability of the system to sustain respective features of time-evolution continuously over  extended periods of time as a result of particular mechanisms associated with the system. 
It is beyond the scope of this paper to constraint the classes of features that are 
suitable for the  identification of the transient regimes that  precisely correspond to life. 
It suffices to assume that the transient regimes of interest are \textit{identifiable operationally}, 
for example, from the special structure of the time-series of concentrations, responses to perturbations, distribution of fluctuations, etc.

The setting depicted in Figure 1 can be interpreted in a different manner. Segment 1-2 can be understood as a long-lived \textit{state} of high free energy. If the activation energy, i.e., the reaction barrier, for such state is sufficiently high, its lifetime is enhanced due to the kinetic hindrance of the relaxation processes. This line of thinking has motivated multiple studies \cite{Ref08,Ref20,Ref21}. Our study proceeds from a different vantage point. The barriers and rate constants are never modified. Instead of the lifetimes of molecules we are concerned with \textit{organization of the processes} that involve molecules. Molecules participate in reactions, reactions form networks, and reaction networks interact with each other. It is the level of the intra- and inter-network interactions that we investigate. Our goal is to identify non-equilibrium modes of time-evolution that are accessible at this level and characterize lifetimes of the encountered transient kinetic regimes.

We deliberately avoid including replication-related events into the diagram because self-replication 
is contingent on the lifetime of the replicating entity. 
If the lifetime is shorter than the replication period, the replicator cannot produce an offspring. 
Such a contingency is accounted for in the models of population dynamics. 
However, lifetimes of replicators in such models are externally defined model parameters. 
In other words, population dynamics and related concepts of the population stability \cite{Ref20,Ref21} can explain selection and fixation of the replicators with appropriately long 
lifetimes, but they do not describe the mechanism of the emergence of long lifetimes. 
Within the concept of sustainability of transient kinetic regimes, death is fundamentally independent 
of the development of self-replicating capability but rather implies some level of complexity 
in the organization of the system, cf. interacting complex reaction networks.

The primary goal of this paper is to illustrate the concept of sustainability of transient kinetic 
regimes that was introduced above. 
To this end, we consider a model extended system - a lattice of compartmentalized chemical oscillators 
under batch and semi-batch conditions. 
In the extended system each compartment is exposed to the effective environment created by 
its immediate and distant neighbors. We show that this leads to a noticeable increase of the sustainability of transient oscillating regime. Specifically, the average lifetimes of oscillations in the compartments increase under batch conditions. As a consequence, the survival function of the oscillations in the extended system is enhanced under semi-batch conditions.
 
Chemical oscillations are a convenient choice of an operationally identifiable transient kinetic regime. Under batch conditions the reactants 
are loaded into the system only once and can be completely consumed. In this case, oscillations have a finite lifetime.
After the oscillations have stopped, the chemical system is still away from the equilibrium, which 
matches the structure of the diagram in Figure~\ref{fig:SchmtcRpsnttn}. 
Multiple compartmentalized oscillators that are allowed to interact with each other, for example, via diffusion of one or several molecular species, form an extended system. 
The most interesting feature of the extended systems is their potential to exhibit collective phenomena, such as spontaneous synchronization of oscillations. This behavior is extensively studied in the classical context following work of Kuramoto \cite{Ref22}, and it was demonstrated to exist 
in the quantum world, too \cite{Ref23}. 
The processes developing in each compartment of the extended system are exposed to the noise produced by the rest of the system. Oscillating kinetic regimes are remarkably sensitive to the intrinsic and extrinsic noise, that can 
serve as i) a trigger of oscillations and ii) a ``control knob'' that modifies parameters of oscillations 
\cite{Ref24,Ref25,Ref26}. 
Spatial confinement has been shown to play role in the development of oscillating regimes \cite{Ref27}. 
Experimental studies of the networks of chemical oscillators, such as test of Turing's theory in morphogenesis 
and observation of chemical differentiation in chemical cells \cite{Ref28}, are facilitated by development of microfluidics.
Cooperative phenomena in extended systems have a capability to evolve as the consequence of the system's growth. 
An example of such evolutionary development is ``dynamical quorum sensing''. 
This term covers a range of phenomena where the population density of compartments controls their
transition to coordinated activities. 
Synchronization of oscillations is a particular case that is studied in colonies of unicellular 
organisms \cite{Ref29} and model chemical systems \cite{Ref30,Ref31}. 
Interestingly, quorum sensing has been implicated as a part of the differentiation mechanisms 
that enable bacterial colonies to adapt to shortage of nutrients in the environment via changing 
developmental program of some cells and inducing death of others \cite{Ref15}.

There is a long standing interest in the phenomenon of oscillation death, that designates loss 
of the oscillations as a result of the interaction between oscillators in an extended system \cite{Ref32,Ref33}. 
The possibility of the extension of the lifetimes of decaying oscillations has not been addressed 
to our knowledge. 
The main reason is that the majority of the studies dedicated to the emergent phenomena in arrays 
of coupled chemical oscillators require oscillations to persist longer than it takes to perform target 
measurements.

Before we proceed to the discussion of the results, we would like to explicitly formulate the considerations 
that motivate our effort and provide the context for the assessment of its outcome:

\begin{itemize}
\item	emphasis on the protection of spatio-temporal organization of processes, i.e., sustainability of kinetic regimes, rather than protection of molecular species via kinetic hindrance of chemical reactions.
\item	exploration of the mechanisms ``built into'' the system, rather than special requirements to the environment.
\item	bias toward the question ``What life is?'', rather ``How did life emerge on Earth?''
\end{itemize}

\section*{Methodology}

We consider the Brusselator system of mass-action kinetic equations \cite{Ref34,Ref35} (Eq. 1)  
as the source of chemical oscillations under batch conditions.

\begin{equation}
\begin{split}
\mathrm{A} & \xrightleftharpoons[k_{-1}]{\,k_{1}\,} \mathrm{X}
\\
\mathrm{B} +\mathrm{X} & \xrightleftharpoons[k_{-2}]{\,k_{2}\,}  \mathrm{Y} + \mathrm{D}
\\
2\mathrm{X} + \mathrm{Y} & \xrightleftharpoons[k_{-3}]{\,k_{3}\,}  3\mathrm{X}
\\
\mathrm{X}  & \xrightleftharpoons[k_{-4}]{\,k_{4}\,}  \mathrm{E}
\end{split}
\end{equation}

The Brusselator is an abstract model that captures the phenomenon of chemical oscillations. 
It includes two reactants, A and B, two intermediates, X and Y, and two products D and E. 
Oscillations are encountered in the concentration profiles of the intermediates X and Y. 
Under open-flow conditions the reactants are available without shortage and the oscillations are 
sustained indefinitely long. 
Under batch conditions the amounts of the reactants are finite, so that the oscillations extinct over finite time as reactants are consumed. Under semi-batch conditions finite amounts of the reactants are added to the system with finite delays. If the reactants are replenished before the oscillations have stopped, there is a possibility that the oscillation lifetime will be extended. Of course, the latter effect is contingent on whether the system remains within the scope of the oscillating regime upon the addition of the reactants. This consideration constrains such parameters as the reactants supply rates, lower and upper bounds on reactants concentrations, etc. In the limit of the vanishing delays the semi-batch regime approaches an open-flow regime, and in the limit of the infinitely long delays it approaches a batch regime. In the rest of the paper we use 
arbitrary units of time and concentration as a consequence of the abstract nature of the kinetic 
model.
\begin{figure*}[htp!]
\begin{tabular}{ccc}
            \includegraphics[width=0.95\textwidth]{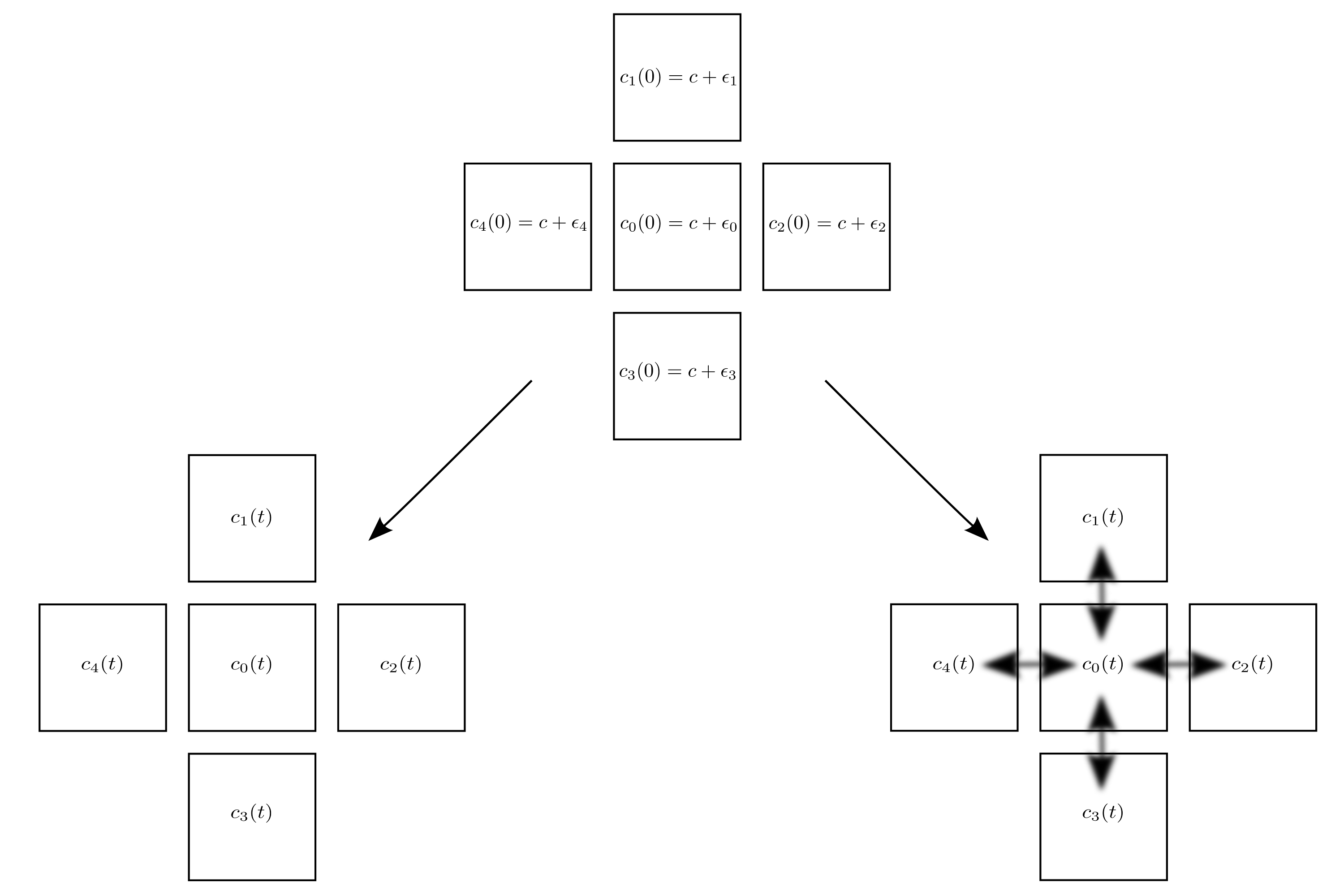} 
\end{tabular}
\caption{The outline of the computational experiment. 
A collection of compartments on a square finite-size lattice is studied. 
Each compartment contains a Brusselator network of reactions (Eq. 1); 
$c_i[c_\mathrm{A},\, c_\mathrm{B},\, c_\mathrm{E},\, c_\mathrm{D},\, c_\mathrm{X},\, c_\mathrm{Y}\,]$ 
designates the concentration vector of the $i$-th compartment, $c_i(0)$ corresponds to the initial state, 
and $c_i(t)$ to some arbitrary time. 
The vector of the initial concentrations ci(0) is formed as a sum of a constant vector 
$c = [c_\mathrm{A}=100,\, c_\mathrm{B}=200,\, c_\mathrm{E}=0,\, c_\mathrm{D}=0,\, 
c_\mathrm{X}=0,\, c_\mathrm{Y}=0]$ and a random vector 
$\epsilon_i = [\epsilon_\mathrm{A}\sim N(\mu,\sigma),\, \epsilon_\mathrm{B} \sim N(\mu,\sigma),\, 
\epsilon_\mathrm{E}=0,\, \epsilon_\mathrm{D}=0,\, \epsilon_\mathrm{X}=0,\, \epsilon_\mathrm{Y}=0]$, 
where $\epsilon_A$  and $\epsilon_B$ are drawn from $N(\mu,\sigma)$, which is a 
normal distribution with mean $\mu=0$ and standard deviation $\sigma=10$ (top middle). We use one set of the inhomogeneous initial concentrations to run kinetic simulations of two versions of the system. In the non-interacting version the compartments are isolated from each other (bottom left). In the interacting version each compartment is diffusively coupled to its four nearest neighbors (bottom right). An ensemble of random realizations of the initial conditions is explored in this manner.}

\label{fig:ExprimentOutline}
\end{figure*}
To illustrate the concept of sustainability of transient kinetic regimes, we construct the 
simplest extended system -- a finite-size square lattice of compartmentalized Brusselators 
(Fig.~\ref{fig:ExprimentOutline}). 
We start with inhomogeneous initial concentrations across the lattice by forming the 
vector of initial concentrations in the $i$-th compartment, $c_i(0)$, in the following way:
\begin{equation}
\begin{split}
c_i(0) &= c + \epsilon_i,
\\
c &= [c_\mathrm{A}=100, \, c_\mathrm{B}=200, \,c_\mathrm{E}=0, 
\,c_\mathrm{D}=0, \,c_\mathrm{X}=0, \,c_\mathrm{Y}=0],
\\
\epsilon&= [\epsilon_\mathrm{A} \sim N(\mu,\sigma),\, \epsilon_\mathrm{B} \sim N(\mu,\sigma),\, 
\epsilon_\mathrm{E} =0,\, \epsilon_\mathrm{D} =0,\, \epsilon_\mathrm{X} =0,\, \epsilon_\mathrm{Y} =0,\,].
\end{split}
\end{equation}
where $N(\mu,\sigma)$ is a normal distribution with mean $\mu=0$ and standard 
deviation $\sigma=10$. We use one set of the inhomogeneous initial concentrations to run kinetic simulations of two versions of the system. In the non-interacting version, the compartments are isolated from each other (bottom left). In the interacting version, each compartment is diffusively coupled to its four nearest neighbors (bottom right). An ensemble of random realizations of the initial conditions is explored in this manner. Further details of the kinetic simulations of the compartmentalized Brusselators are provided in Supporting Information. 

The model in the described form has a sizable parameter space. 
We will discuss a single combination of the parameters in order to accomplish the main goal of our study -- to 
illustrate the concept of sustainability of transient kinetic regimes. 
Specifically, we limit simulations to a 30$\times$30 square lattice with 5000 random realizations 
of the inhomogeneous initial conditions. 
Initial concentrations of A and B are drawn from a normal distribution with mean of 100 and 200 
units, respectively, both with standard deviation of 10. 
Compartments are coupled via intermediate Y with a homogeneous coupling constant $k_{c_\mathrm{Y}}=0.08$. 
A detailed investigation of the parameter space and characterization of all possible kinetic regimes 
is left for future studies.

We use the following procedure to determine the oscillation lifetimes in the compartments. First, the 
moving-average is constructed for the concentration profile of the oscillating component. Second, the points of intersection between the original and averaged time-series are found. Finally, the first 
and the last intersection points are taken as the times of the oscillation onset and extinction, so that the oscillation lifetime can be obtained. We truncate the time-series if separation between two consecutive intersection points becomes too large. This step helps to avoid artifacts associated with slowly developing non-monotonic concentration 
changes (see Fig.~S1 in Supporting Information for details). 
The oscillation lifetimes in the $i$-th compartment of the diffusively coupled (interacting) system is designated as  $\lambda_{c_i}$; the oscillation lifetime in the same compartment of the uncoupled (non-interacting) system is $\lambda_{u_i}$. We define the relative lifetime enhancement for the $i$-th compartment $\gamma_i$ as a ratio of these lifetimes:
\begin{equation}
\gamma_i = \lambda_{c_i}/\lambda_{u_i}.
\end{equation}

Oscillators with infinite relative lifetime, i.e., those that do not exhibit oscillations in the uncoupled 
system, are assigned $\gamma=0$. 
If the average period of the oscillations in the coupled system increases more than 2-fold relative 
to the uncoupled case, the relative lifetimes are treated as unreliable and such oscillators are also 
assigned $\gamma=0$.

\section*{Results and Discussion}

First, we note that the lifetimes of the compartmentalized oscillations depend on the coupling strength 
between the compartments. Relative lifetime $\gamma$ is trivially 1 in the limit of vanishing diffusion constant $k_{c_\mathrm{Y}}$. It goes to 0 as the diffusion constant becomes much larger than the rate constants of the Brusselator model. The latter behavior fits into the concept of the oscillation death \cite{Ref32,Ref33}. It should be encountered in the models with dimensionless rate constants and coupling constants, such as the  one considered in the present study. Such behavior in the limit of strong diffusive coupling is not transferable to the models that involve dimensional constants with different dimensionalities.

Next, we compare the raw data that describe oscillation lifetimes in the non-interacting and 
interacting (coupling constant $k_{c_\mathrm{Y}}=0.08$) systems. 
The histograms of the lifetimes of each individual compartment over the ensemble of the simulated lattices are shown in the left panel of Figure~3. 
The histogram of the non-interacting system has a single sharp feature at 100 
time units. The width of the feature is due to the variance in the initial concentrations of the reactants between compartments (see Methodology). 
In the case of the interacting system the histogram has two broad features at 160 and 320 
time units. We used the same ensemble of the initial concentrations to run kinetic simulations of the interacting and non-interacting systems. Therefore, the higher count of longer lifetimes is the consequence of the diffusive coupling between compartments. The central panel of Figure~3 shows the histogram of the relative lifetimes $\gamma$ computed 
according to Eq.~2 for each individual compartment over the simulation ensemble. 
The peak of the histogram corresponds to $\gamma=1.5$ and a shoulder is formed at $\gamma = 3$.

\begin{figure*}[htp!]
\begin{tabular}{ccc}
            \includegraphics[width=0.95\textwidth]{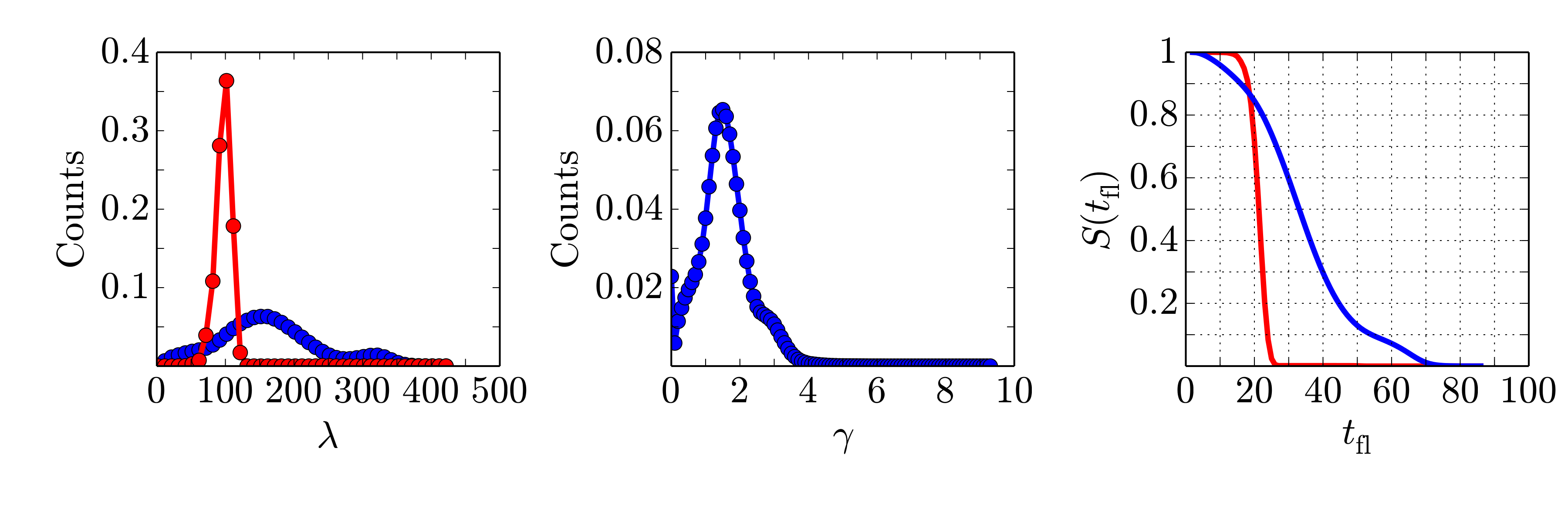}                                      
\end{tabular}
\caption{Left panel: histograms of the oscillation lifetimes of each individual compartment over an ensemble of non-interacting 
(red, $\lambda_u$) and interacting (blue, $\lambda_c$) lattices. 
Central panel: histogram of the relative lifetimes $\gamma$ for each individual compartment over the simulation ensemble. 
Right panel: survival function of the transient oscillating regime $S(t_{fl})=P(\lambda>t_{fl})$, where $t_{fl}$ 
is feeding lag and $\lambda$ is oscillation lifetime. 
$S(t_{fl})$ is the fraction of compartments that can sustain oscillations between multiple feedings 
that occur with a constant feeding lag $t_{fl}$. 
Red curve corresponds to the non-interacting system, blue curve -- to the interacting.}
   \label{fig:LifeTimes}
\end{figure*}
So far we were concerned with the outcomes of the simulations under batch conditions. Now we will investigate how differences between interacting and non-interacting systems play out under semi-batch conditions. In this case, reactants A and B can be added to the compartments multiple times with some delay. We will refer to this delay as a "feeding lag" $t_{\mathrm{fl}}$. For simplicity, we will assume that the delivered amounts of the reactants effectively reset their concentrations to the values consistent with the procedure that was used to generate initial conditions for the kinetic simulations under batch regime (see Eq. 2 and Methodology), reactants are delivered to all compartments simultaneously, i.e., a single feeding lag value applies to the entire lattice, and the feeding lag is constant. This choice of the model enables us to describe semi-batch regime using the information obtained from batch simulations. If the lifetime of oscillations in the $i$-th compartment, $\lambda_i$, is longer than the lag $t_{\mathrm{fl}}$, the reactants are replenished before the oscillations die out, so that the transient regime is sustained without failure; otherwise, the transient regime is lost. Therefore, we can describe the behavior of the interacting and non-interacting systems under semi-batch conditions in terms of the survival function of the oscillations in the lattice compartments. We use the following definition of the survival function $S(t_{\mathrm{fl}})$:
\begin{equation}
\begin{split}
S_i(t_{\mathrm{fl}}) &= P(\lambda_i>t_{\mathrm{fl}}),
\\
P(\lambda_i>t_{\mathrm{fl}}) &= \int\limits_{t_{\mathrm{fl}}}^{\infty} \mathrm{d}\lambda_i f(\lambda_i).
\end{split}
\end{equation}

Here, $S_i(t_{\mathrm{fl}})$ is the survival function for the $i$-th compartment. It is given by the probability $P(\lambda_i>t_{\mathrm{fl}})$ that the oscillations in the compartment last longer than some value $t_{\mathrm{fl}}$. $f(\lambda_i)$ is the probability distribution of the oscillation lifetimes in the respective compartment. Using frequencies of raw oscillation lifetimes (Fig.~\ref{fig:LifeTimes}, left panel) as proxies for $f(\lambda)$, we obtain survival functions for the interacting and non-interacting versions of the system under semi-batch regime (Fig.~\ref{fig:LifeTimes}, right panel). As one would expect, a very narrow distribution of the oscillation lifetimes in the non-interacting system (Fig.~\ref{fig:LifeTimes}, left panel, red curve) causes a very fast fall-off of the survival function. If the feeding lag exceeds 25 time units, there is essentially zero probability of the survival of the transient oscillating regime on the lattice (Fig.~\ref{fig:LifeTimes}, right panel, red curve). The lifetime distribution in the interacting system is broader and its peak is shifted toward higher values (Fig.~\ref{fig:LifeTimes}, left panel, blue curve); the lifetimes are typically enhanced by a factor of 1.5 relative to the non-interacting case (Fig.~\ref{fig:LifeTimes}, central panel). As a consequence, the survival function for the interacting system indicates much higher sustainability of the transient kinetic regime. For example, feeding lag of 25 time units corresponds to the survival of oscillations in $~70\%$ of the compartments. 
Up to $10\%$ of the compartment will sustain the transient regime with the feeding lag up to 54 time units. The complete loss of the transient regime will occur if the lag exceeds 74 time units, which extends the limits of sustainability by factor of 3 in comparison to the uncoupled system. Of course, a more complicated model of the semi-batch regime can be considered, that includes 
randomization of the feeding lag values between compartments and between feeding events. 
As long as the interacting system has longer lifetimes per one feeding event, the qualitative result, i.e., higher sustainability of the transient regime in the interacting system, will hold.

\begin{figure*}[htp!]
            \includegraphics[width=0.95\textwidth]{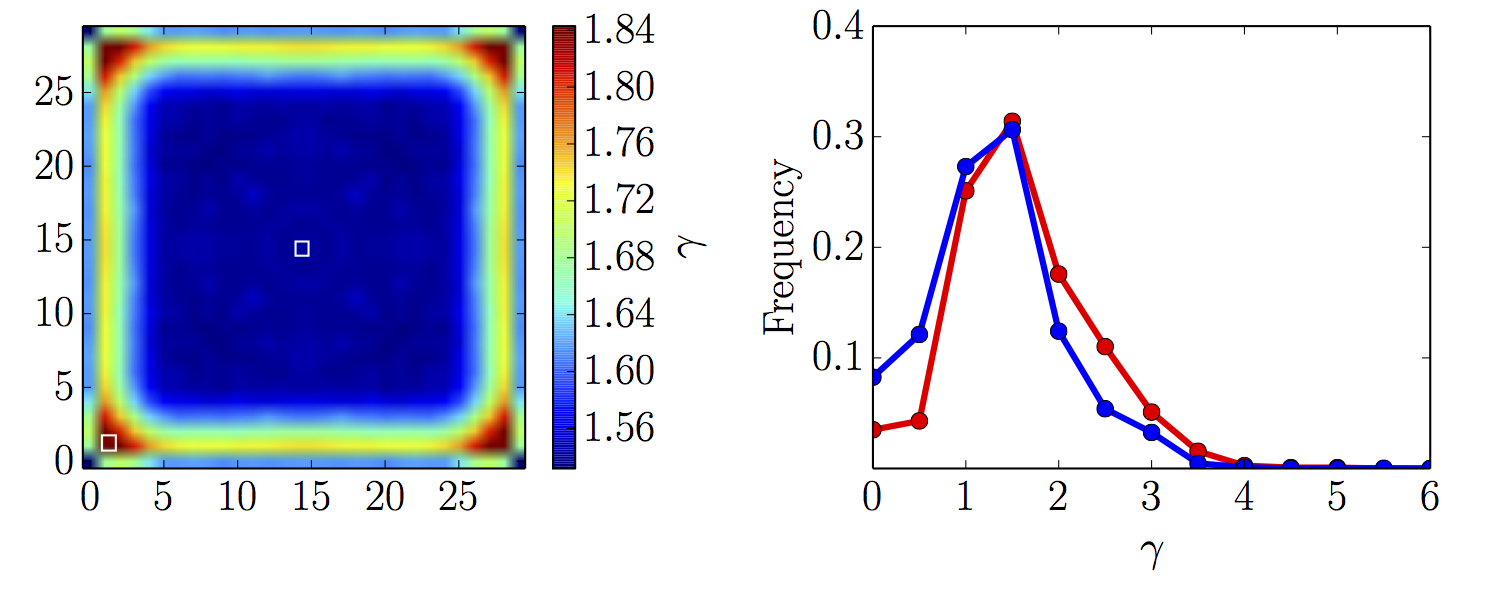} 
\caption{Site-specific distribution of the relative lifetimes. 
Left panel: spatial distribution of the relative lifetime $\gamma$ on the lattice after averaging over the ensemble 
of realizations of the initial conditions; the distribution is symmetrized. 
Right panel: histogram of the frequency distributions of relative lifetimes for the near-corner site (red) 
and central site (blue) obtained from the ensemble of realizations of the initial conditions; both sites are 
marked with white squares on the left panel.}
\label{fig:LifeTimeDistrib}
\end{figure*}

All the results discussed in this paper are obtained using a finite-size model. It is instructive to discern the spatial structure of the distribution of the relative lifetimes $\gamma$ (Fig.~\ref{fig:LifeTimes}, 
central panel) in order to understand the role of the edges. 
We average relative lifetimes for each compartment over the ensemble of the model realizations;  
the obtained compartment-specific values are further averaged between the lattice sites related 
by symmetry on square lattice. 
The symmetrized distribution of the averaged site-specific relative lifetimes is shown in the right panel 
of Figure~\ref{fig:LifeTimeDistrib}. 
The relative lifetime of the transient regime in the compartments in the vicinity of the edges is higher 
than in the ``bulk'' region; it reaches the highest values close to the lattice corners. Right panel of Figure~\ref{fig:LifeTimeDistrib} shows histograms of the frequencies of the relative 
lifetimes for the near-corner site and the central site of the lattice over the simulation ensemble. 

The histograms have  very similar shapes with peaks at $\gamma\approx1.5$. 
The histogram of the near-corner site shows higher frequencies of longer relative lifetimes including 
a shoulder at $\gamma\approx3.0$, that explains the shoulder on the histogram in the central panel 
of Figure~3. In contrast, the histogram of the central site shows enhanced frequencies of lower relative lifetimes. 
This explains the difference in the averaged values for the respective regions of the lattice. 
From the structural point of view, the edges and the corners of the lattice are 1- and 0-dimensional defects, 
respectively. 
A typical simulation strategy might aim to remove or minimize such effects in order to understand the behavior 
of ``bulk''. 
However, the goal of our study is to discuss the factors that facilitate emergence of the sustainable transient 
kinetic regimes. 
The role of the lattice defects in this regard is very interesting. It hints that a less trivial spatial structure 
of the extended system, such as a network of small lattices that effectively mimic 0- and 1-dimensional defects, 
might lead to a much stronger improvement of the sustainability. 

\begin{figure*}[htp!]
\includegraphics[width=0.95\textwidth]{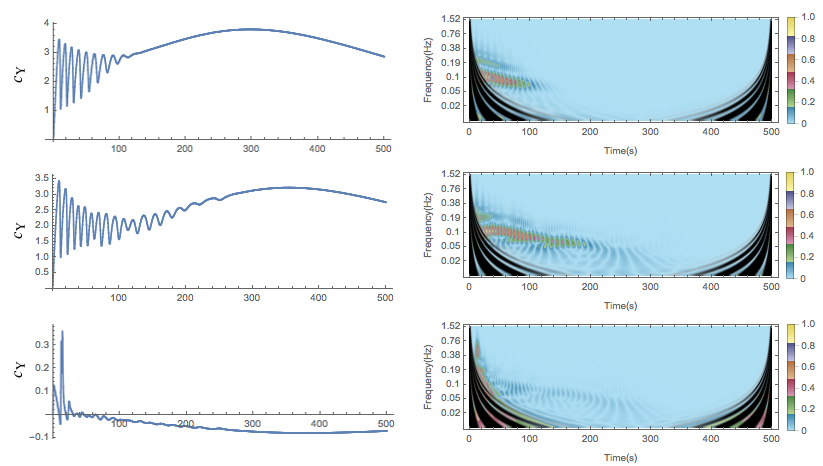}
\caption{Left panels: time series of the concentration of component Y, $c_{\mathrm{Y}}(t)$, 
in a representative compartment of the non-interacting system (top), $c_{\mathrm{Y}}(t)$ in the same compartment of the interacting system (center), and time series of the noise produced by the effective environment of the respective compartment in the interacting system (bottom). Right panels: scalograms of the continuous wavelet transform of the time series in the left panels. The scalogram encodes the wavelet coefficients using a color function represented by the color bar. The color function is scaled, so that its arguments lie in the range 0 to 1. Large coefficients correspond to the scales that contribute the most to the time-series. All three scalograms have their main features at the scale $\sim 0.1$~Hz. This feature is long-lived on the scalograms that correspond to the interacting systems. Black color masks cone of influence (COI). COI shows the regions of the scalogram that are strongly affected by the endpoints of the time-series. 
}
\label{fig:Wavelets}
\end{figure*}

Finally, we can discuss some aspects of the mechanism that ensures the enhancement of the oscillation 
lifetimes in the system of coupled compartments. 
In the non-interacting case, i.e., uncoupled lattice, the interactions of any compartment with the environment are 
limited to the introduction of the reactants A and B into the compartment. However, in the interacting system the notion of the environment changes. 
In addition to the environment that serves as the source of reactants, any given compartment can be considered as embedded in the effective environment that represents the rest of the lattice and interacts with the compartment via diffusion of one or several components. 
Therefore, we will treat the interacting system in the spirit of mean-field approximation and replace the effective environment of a single compartment by an effective field (noise) that drives the single compartment (system).
The strongest contributors to the noise are the nearest neighbors of the compartment on the lattice. This consideration simplifies derivation of the analytical expression for the noise as a function of time (see Eq.~S15 in SI). 
Let us pick a representative compartment that shows enhancement of the oscillation lifetime in the interacting system. 
The time-series of interest (Fig.~\ref{fig:Wavelets}, left panel) are the concentration time-series of the component Y, $c_{\mathrm{Y}}(t)$, in this compartment in the non-interacting system, $c_{\mathrm{Y}}(t)$ in the interacting system, and the time-series of the noise due to the effective environment of this compartment in the interacting system (Eq.~S15 in SI). 
The dominant modes of these three time-series can be extracted using continuous wavelet transform. The right panel of Figure~\ref{fig:Wavelets} shows corresponding wavelet scalograms. All three scalograms have a feature $\sim 0.1$~Hz. This feature persists for $\sim100$ time units in the non-interacting case. 
In the interacting case, it extends to $\sim 200$ time units in both $c_{\mathrm{Y}}$ and the noise time-series. 
The cross-spectrum wavelet analysis of $c_{\mathrm{Y}}$ and the noise time-series in the interacting system (see FIg. S2 in SI) indicates that the respective dominant modes remain coherent over the time of the sustained oscillations. 

We interpret these results as a fingerprint of causality between the time-evolution of the effective environment and a single compartment, i.e., the system. The noise produced by the effective environment acts on the damped non-linear chemical oscillations in the system as an external driving force. The effective environment is not completely controlled by the proceesses in the system, yet it is constitutionally related to the system. This leads to the resonance-like relationships between the respective signals. Tentatively, we identify these relationships as a form of stochastic resonance \cite{SR,SR0}.  This phenomenon has been explored in the broad context of chemcial oscillations \cite{Ref25,SR1,SR2}. 
In general, it refers to the modification, typically enhancement, of the system performance in the presence of noise. We note that the definitions and effects under the umbrella of the term "stochastic resonance" are diverse and not without controversies \cite{SR0}. Our particular case fits into the "bona fide" concept of the resonance because the system and the noise have matching dominant modes in their time-series and the amplitude of the decaying oscillations in the system is enhanced due to the noise. 
As in the case of linear oscillators, inter-oscillator coupling and intrinsic noise have the effect of changing the spectrum of the local oscillators. In particular, for some oscillators the central spectral peak shifts towards higher frequencies (see, e.g., Ref. 26). This means that for the same noise strength, the central frequency is more robust to that noise and therefore has a longer lifetime.
We reserve detailed technical analysis of the dynamics of the extended system in terms of the phase space structure, such as types of attractors and bifurcations, for the future studies.

Overall, effective environment emerging in our model is different from the regular thermal bath. 
It serves as a sink of energy for the system and a source of the driving force that becomes stochastic in the limit of the infinite number of compartments yet remains constitutionally related to the processes that develop in the system.

The central motivation of the presented study is to introduce the concept of sustainability 
of transient kinetic regimes and to illustrate it via a numerical study of an abstract model. 
This model, however, can be taken as a basis for the experimental studies to validate 
the predicted behavior. 
It can be implemented literally, as a network of diffusively coupled compartmentalized chemical 
oscillators, such as Belousov-Zhabotinsky reactions \cite{Ref36} or biochemical networks \cite{Ref37,Ref38}. 
Compartmentalization can be achieved using vesicles \cite{Ref28}, loaded catalytic particles \cite{Ref30} or even 
porous medium under percolation conditions. 
Growth of the density of compartments is the most obvious mechanism that renders such systems 
evolvable, cf. the studies of dynamic quorum sensing \cite{Ref29,Ref30,Ref31}; another evolutionary factor is 
development of the spatial structure of the inter-compartment interactions. 
It can have multiple forms ranging from the regular lattices to random networks with static or 
dynamic coupling. 
A very different implementation comes to mind that is based on development of a network of coupled 
molecular vibrations, such as vibrations in polymers, where evolution of the system includes growth 
of the polymer chain and increase of the polymer concentration.

The choice of the oscillating transient regime was motivated by the operational simplicity of the 
analysis. 
At this point, there is no evidence of chemical oscillations in  protometabolic chemistry \cite{Ref03}; 
they could be envisioned, however, in the context of the emergence of informational and catalytic 
polymers and development of the precursors of gene regulatory networks. 
The nature of the transient kinetic regimes in prebiological chemical systems is, therefore, a question 
that needs further investigation. 
The answer will depend critically on what kind of systems and chemical transformations are considered. 
Protometabolic networks \cite{Ref03} are the obvious candidates for the analysis, that can proceed along the 
lines explored in other branches of chemistry \cite{Ref39}. 

Having discussed behavior of the compartmentalized oscillators on a lattice, we can return to the 
conjecture proposed in the beginning of the paper: ``Death comes into play in prebiological world 
as a point of failure of the intrinsic mechanisms that ensure sustainability of the transient kinetic regimes''. 
Let us enumerate implicit and explicit parameters of the studied model and evaluate their contribution 
to the sustainability of a transient kinetic regime. 
One group of implicit parameters is the lifetimes of molecules that participate in the reactions. 
These lifetimes depend on the barriers that ultimately determine the rate constants of the reactions 
(Eq.~1). 
The values of the rate constants of the Brusselator have to be related via some ratios in 
order for oscillations to be possible \cite{Ref34}. 
Therefore, emergence of any kinetic regime is contingent on a specific hierarchy of molecular lifetimes which is an intrinsic feature. 
Any factor that changes this hierarchy, such as temperature, catalysis, or solvent, will change the viability 
and baseline characteristics of the kinetic regimes that can be observed. 
These factors are, however, extrinsic. 
Another group of implicit parameters covers structural properties of the compartments. 
We assumed that the structural integrity of the compartments is maintained over times 
that exceed the lifetime of the transient kinetic regime of interest. 
We also assumed that the composition and structure of the compartment wall do not change 
with time so that the diffusion constant is time-independent. 
These factors are extrinsic, but they can be modified due to the interactions of the compartment 
walls with reaction components. 
Such interactions can be easily introduced into the abstract model, but it is more important to 
find an actual realization of a system where they can exist. 
Finally, the spatial structure of the extended system was treated as frozen, which is a simplification. 
This factor is also extrinsic, but it can develop sensitivity to the progress of the compartmentalized 
reactions if the compartment structure is modified by the reaction intermediates or products. 
For example, collapse of the compartments, their ``swelling'' \cite{Ref28}, or accumulation of charged 
species in the compartment walls can factor into the evolution of the spatial structure of the 
extended system. 
Overall, formation of an extended system and emergence of the effective environment serves as a bridge between extrinsic and intrinsic factors. Effective environment 
becomes a part of the mechanism that increases sustainability of the transient kinetic regime. 
Such mechanism will fail if the interactions within the extended system are compromised, e.g., 
due to compartment disintegration and/or degradation of the coupling between compartments. 

The most important and the least model-dependent conclusion of our work concerns relationships 
between the system and its environment as a factor that determines the fate of the non-equilibrium 
systems. 
Presence of a dissipative environment is the single reason that leads to the onset of non-equilibrium 
\cite{Ref40}regimes and emergence of structure \cite{Ref11}. 
However, there are no known laws of nature, that grant that some such regimes will ``learn'' to 
maintain the appropriate external conditions and environments thus prolonging their own existence. 
Extended systems offer a way out of this conundrum. 
In our model formation of the extended system facilitates emergence of the effective 
environment interacting with a single compartment. In this system-environment partitioning the environment remains constitutionally related to the system. This leads to the increase of the lifetime of a non-trivial transient regime within the system, at least within the boundaries of the model applicability. 
The importance of tailoring the dissipative environments is realized in the area of quantum 
applications \cite{Ref40} where particular dissipative environments are capable of supporting extremely 
long-lived oscillations \cite{Ref41}. In this context, there have been developed strategies to characterize experimentally the spectral properties of the environments \cite{Ref42}. 
The emergence and evolution of the dissipative environments should become a focus of the prebiological chemistry along with the emergence and evolution of prebiochemical systems.

%\noindent LaTeX formats citations and references automatically using the bibliography records in your .bib file, which you can edit via the project menu. Use the cite command for an inline citation, e.g.  \cite{Figueredo:2009dg}.

\section*{Acknowledgements}

This work was supported by a grant from the Simons Foundation (SCOL 291937 to D.Y.Z.) 
L.A.P acknowledges support from Comit\'e para el Desarrollo de la Investigaci\'on-CODI-of Universidad de
Antioquia, Colombia under the Estrategia de Sostenibilidad 2014-2015, by the
Colombian Institute for the Science and Technology Development-COLCIENCIAS-under the Contract 
No.~111556934912. Authors thank Dr. J. McClean for the inspiring discussions.

\section*{Author contributions statement}

D.Y.Z. and L.A.P. conceived the project, developed methodology, carried out computations and wrote the manuscript.

\section*{Additional information}

The authors declare no competing financial interests.

\newpage


\section*{Supplementary material}

# Supporting Information for:

# Sustainability of Transient Kinetic Regimes and Origins of Death

Dmitry Yu. Zubarev[1] and Leonardo A. Pachón[2]

[1]*Department of Chemistry and Chemical Biology,*

*Harvard University, Cambridge, MA 02138 USA.*

[2]*Grupo de Física Atómica y Molecular, Instituto de Física,*

*Facultad de Ciencias Exactas y Naturales,*

*Universidad de Antioquia UdeA; Calle 70 No. 52-21, Medellín, Colombia.*

(Dated: October 10, 2015)

# I. KINETIC SIMULATION OF COMPARTMENTALIZED COUPLED OSCILLATORS

$$k_1 = 0.007;$$
$$k_{-1} = 0.0001;$$
$$k_2 = 0.01;$$
$$k_{-2} = 0.0001;$$
$$k_3 = 1;$$
$$k_{-3} = 0.0001;$$
$$k_4 = 1;$$
$$k_{-4} = 0.0001;$$

Numerical integration carried out in MatLab R2014.a. using standard numerical integrator "ode45" with integration step 0.005 over 100000 steps.

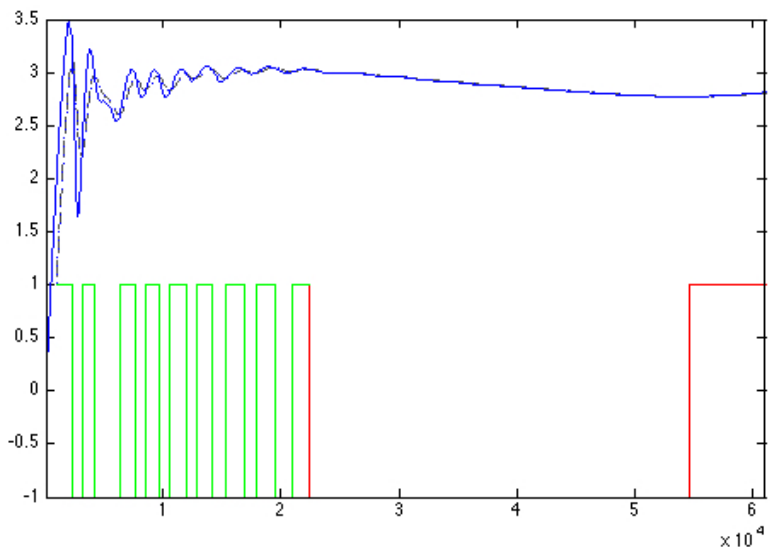

FIG. 1. Operational estimation of the lifetime of the oscillatory regime. The solid line is the concentration time-series for the intermediate Y obtained from the kinetic simulations; the dashed grey line is its moving-average.

We evaluate the sign function of the difference between the raw and the averaged time-series:

$$\text{sign}(a - b) = 1, a > b,$$
$$\text{sign}(a - b) = -1, a < b. \quad (1)$$

Oscillation is considered to persist for as long as the sign function changes the value. We introduce a threshold on the ratio between two consecutive intervals for the change of sign in order to eliminate contamination due to slow non-monotonous concentration changes. In this example, the solid red line and solid green line are plots of the sign function. The former picks up contribution from a a slow non-monotonous concentration change and yields longer estimates of the oscillation lifetimes. Instead, we use the truncated data (green line) to obtain lifetime values.

## II. EFFECTIVE ENVIRONMENT PERSPECTIVE

Derivation of the expression for noise in the limit of the nearest neighbor contribution:

$$\dot{A} = -k_1 A + k_{-1} X, \quad (2)$$
$$\dot{B} = -k_2 B + k_{-2} DY, \quad (3)$$
$$\dot{D} = k_2 BX - k_{-2} DY, \quad (4)$$
$$\dot{E} = k_4 X - k_{-4} E, \quad (5)$$
$$\dot{X} = k_1 A - k_{-1} X - k_{-2} DY + k_3 X^2 Y - k_{-3} X^3 - k_4 X + k_{-4} E, \quad (6)$$
$$\dot{Y} = k_2 BX - k_{-2} DY - k_3 X^2 Y + k_3 X^3 + \sum_n c_n (Y_n - Y), \quad (7)$$

$$\dot{A}_n = -k_1 A_n + k_{-1} X_n, \tag{8}$$

$$\dot{B}_n = -k_2 B_n + k_{-2} D_n Y_n, \tag{9}$$

$$\dot{D}_n = k_2 B_n X_n - k_{-2} D_n Y_n, \tag{10}$$

$$\dot{E}_n = k_4 X_n - k_{-4} E_n, \tag{11}$$

$$\dot{X}_n = k_1 A_n - k_{-1} X_n - k_{-2} D_n Y_n + k_3 X_n^2 Y_n - k_{-3} X_n^3 - k_4 X_n + k_{-4} E_n, \tag{12}$$

$$\dot{Y}_n = k_2 B_n X_n - k_{-2} D_n Y_n - k_3 X_n^2 Y_n + k_3 X_n^3 - c_n(Y - Y_n), \tag{13}$$

$$\ddot{Y} - k_2 \dot{B} X - k_2 B \dot{X} + k_{-2} \dot{D} Y + k_{-2} D \dot{Y} + 2 k_3 X \dot{X} Y + k_3 X^2 \dot{Y} + 3 k_3 X^2 \dot{X} + \dot{Y} \sum_n c_n^2 - Y \sum_n c_n^2$$
$$= \sum_n c_n \left( k_2 B_n X_n - k_{-2} D_n Y_n - k_3 X_n^2 Y_n + k_3 X_n^3 - c_n Y_n \right), \tag{14}$$

$$\ddot{Y} - k_2 \dot{B} X - k_2 B \dot{X} + \left[ k_{-2} \dot{D} + 2 k_3 X \dot{X} - \sum_n c_n^2 \right] Y + \left[ k_{-2} D + k_3 X^2 + \sum_n c_n^2 \right] \dot{Y} + 3 k_3 X^2 \dot{X}$$
$$= \sum_n c_n \left( k_2 B_n X_n - k_{-2} D_n Y_n - k_3 X_n^2 Y_n + k_3 X_n^3 - c_n Y_n \right), \tag{15}$$

Because the term at the right-hand side does not depend functionally on $Y$, it can be assumed to be a given function of time $\xi(t)$.

$$\ddot{Y} - k_2 \dot{B} X - k_2 B \dot{X} + \left[ k_{-2} \dot{D} + 2 k_3 X \dot{X} - \sum_n c_n^2 \right] Y$$
$$+ \left[ k_{-2} D + k_3 X^2 + \sum_n c_n^2 \right] \dot{Y} + 3 k_3 X^2 \dot{X} = \xi(t) \tag{16}$$

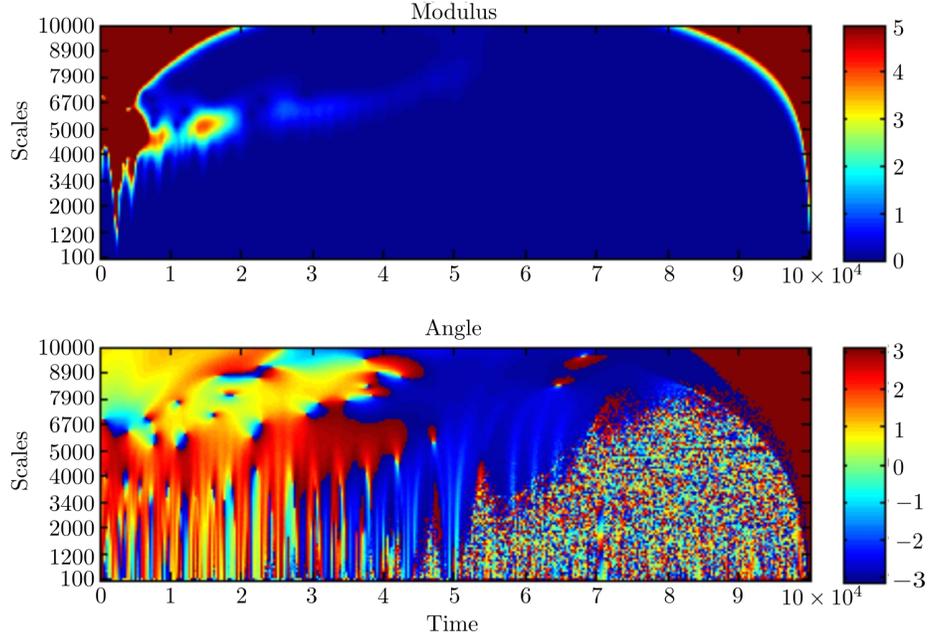

FIG. 2. Wavelet cross spectrum for the time series of $c_Y$ oscillations in the compartment and its effective environment noise (Fig. 5 in the main text). The top panel shows the magnitude of the wavelet cross spectrum; the bottom panel shows the wavelet coherence. Both signals have significant contribution between scales 5600 and 6700 over the interval [0;50000].




\begin{thebibliography}{10}
\expandafter\ifx\csname url\endcsname\relax
  \def\url#1{\texttt{#1}}\fi
\expandafter\ifx\csname urlprefix\endcsname\relax\def\urlprefix{URL }\fi
\providecommand{\bibinfo}[2]{#2}
\providecommand{\eprint}[2][]{\url{#2}}

\bibitem{Ref01}
\bibinfo{author}{H{\"u}gler, M.} \& \bibinfo{author}{Sievert, S.~M.}
\newblock \bibinfo{title}{Beyond the Calvin cycle: Autotrophic carbon fixation
  in the ocean}.
\newblock \emph{\bibinfo{journal}{Annu. Rev. Mar. Sci.}}
  \textbf{\bibinfo{volume}{3}}, \bibinfo{pages}{261--289}
  (\bibinfo{year}{2011}).

\bibitem{Ref02}
\bibinfo{author}{Miller, S.~L.} \& \bibinfo{author}{Urey, H.~C.}
\newblock \bibinfo{title}{Organic compound synthesis on the primitive Earth}. 
\newblock  \emph{\bibinfo{journal}{Science}}
 \textbf{\bibinfo{volume}{130}},
  \bibinfo{pages}{245--251} (\bibinfo{year}{1959}).

\bibitem{Ref03}
\bibinfo{author}{Patel, B.~H.}, \bibinfo{author}{Percivalle, C.},
  \bibinfo{author}{Ritson, D.~J.}, \bibinfo{author}{Duffy, C.~D.} \&
  \bibinfo{author}{Sutherland, J.~D.}
\newblock \bibinfo{title}{Common origins of rna, protein and lipid precursors
  in a cyanosulfidic protometabolism}.
\newblock \emph{\bibinfo{journal}{Nat. Chem.}}
  \textbf{\bibinfo{volume}{7}}, \bibinfo{pages}{301--307}
  (\bibinfo{year}{2015}).

\bibitem{Ref04}
\bibinfo{author}{Seager, S.} \& \bibinfo{author}{Bains, W.}
\newblock \bibinfo{title}{The search for signs of life on exoplanets at the
  interface of chemistry and planetary science}.
\newblock \emph{\bibinfo{journal}{Sci. Adv.}} 
  \textbf{\bibinfo{volume}{1}}, \bibinfo{pages}{e1500047} (\bibinfo{year}{2015}).

\bibitem{Ref05}
\bibinfo{author}{Eigen, M.} \& \bibinfo{author}{Schuster, P.}
\bibinfo{title}{Hypercycle - principle of natural self-organization. Part A: Emergence of the Hypercycle}.
\newblock \emph{\bibinfo{journal}{Naturwissenschaften}}
  \textbf{\bibinfo{volume}{64}}, \bibinfo{pages}{7--41} (\bibinfo{year}{1977}).

\bibitem{Ref06}
\bibinfo{author}{Chen, I.~A.} \& \bibinfo{author}{Nowak, M.~A.}
\newblock \bibinfo{title}{From prelife to life: How chemical kinetics become
  evolutionary dynamics}.
\newblock \emph{\bibinfo{journal}{Acc. Chem. Res.}}
  \textbf{\bibinfo{volume}{45}}, \bibinfo{pages}{2088--2096}
  (\bibinfo{year}{2012}).

\bibitem{Ref07}
\bibinfo{author}{Smith, E.} \& \bibinfo{author}{Morowitz, H.~J.}
\newblock \bibinfo{title}{Universality in intermediary metabolism}.
\newblock \emph{\bibinfo{journal}{Proc. Natl. Acad. Sci. USA}}
  \textbf{\bibinfo{volume}{101}}, \bibinfo{pages}{13168--13173}
  (\bibinfo{year}{2004}).

\bibitem{Ref08}
\bibinfo{author}{England, J.~L.}
\newblock \bibinfo{title}{Statistical physics of self-replication}.
\newblock \emph{\bibinfo{journal}{J. Chem. Phys.}}
  \textbf{\bibinfo{volume}{139}}, \bibinfo{pages}{121923} (\bibinfo{year}{2013}).

\bibitem{Ref09}
\bibinfo{author}{Segr\'e, D.}, \bibinfo{author}{Ben-Eli, D.} \&
  \bibinfo{author}{Lancet, D.}
\newblock \bibinfo{title}{Compositional genomes: Prebiotic information transfer
  in mutually catalytic noncovalent assemblies} 
\newblock \emph{\bibinfo{journal}{Proc. Natl. Acad. Sci. USA}}  
  \textbf{\bibinfo{volume}{97}},
  \bibinfo{pages}{4112--4117} (\bibinfo{year}{2000}).

\bibitem{Ref10}
\bibinfo{author}{Glansdorff, P.} \& \bibinfo{author}{Prigogine, I.} 
\newblock \bibinfo{title}{On general evolution criterion in macroscopic physics}. \newblock \emph{\bibinfo{journal}{Physica}}
  \textbf{\bibinfo{volume}{30}}, \bibinfo{pages}{351-374} (\bibinfo{year}{1964}).

\bibitem{Ref11}
\bibinfo{author}{Prigogine, L.}
\newblock \bibinfo{title}{Time, structure, and fluctuations}.
\newblock \emph{\bibinfo{journal}{Science}} \textbf{\bibinfo{volume}{201}},
  \bibinfo{pages}{777--785} (\bibinfo{year}{1978}).

\bibitem{Ref12}
\bibinfo{author}{Lucia, U.}
\newblock \bibinfo{title}{Stationary open systems: A brief review on contemporary theories on irreversibility}.
\newblock \emph{\bibinfo{journal}{Physica A}} \textbf{\bibinfo{volume}{392}},
  \bibinfo{pages}{1051--1062} (\bibinfo{year}{2013})

\bibitem{Ref13}
\bibinfo{author}{Pulselli, R. M.}, \bibinfo{author}{Simoncini, E.} \& \bibinfo{author}{Tiezzi, E.}
\newblock \bibinfo{title}{Self-organization in dissipative structures: A thermodynamic
  theory for the emergence of prebiotic cells and their epigenetic evolution}.
\newblock \emph{\bibinfo{journal}{BioSys.}} \textbf{\bibinfo{volume}{96}},
  \bibinfo{pages}{237 -- 241} (\bibinfo{year}{2009}).

\bibitem{Ref14}
\bibinfo{author}{Schr{\"o}dinger, E.}
\newblock \emph{\bibinfo{title}{What is life?}} (\bibinfo{publisher}{Cambridge University Press, Cambridge, UK, 1944}).

\bibitem{Ref15}
\bibinfo{author}{Ameisen, J.~C.}
\newblock \bibinfo{title}{On the origin, evolution, and nature of programmed
  cell death: A timeline of four billion years}.
\newblock \emph{\bibinfo{journal}{Cell Death Differ.}}
  \textbf{\bibinfo{volume}{9}}, \bibinfo{pages}{367--393}
  (\bibinfo{year}{2002}).

\bibitem{Ref16}
\bibinfo{author}{Allocati, N.}, \bibinfo{author}{Masulli, M.},
  \bibinfo{author}{Di~Ilio, C.} \& \bibinfo{author}{De~Laurenzi, V.}
\newblock \bibinfo{title}{Die for the community: An overview of programmed cell
  death in bacteria}.
\newblock \emph{\bibinfo{journal}{Cell Death Dis.}}
  \textbf{\bibinfo{volume}{6}}, \bibinfo{pages}{e1609} (\bibinfo{year}{2015}).

\bibitem{Ref17}
\bibinfo{author}{Olson, H.}
\newblock \emph{\bibinfo{title}{Dynamical analogies.}} (\bibinfo{publisher}{Van Nostrand, 1958}).

\bibitem{Ref18}
\bibinfo{author}{Colyvan, B.~M.} \& \bibinfo{author}{Ginzburg, L.~R.}
\newblock \bibinfo{title}{Analogical thinking in ecology: Looking beyond
  disciplinary boundaries}.
\newblock \emph{\bibinfo{journal}{Q. Rev. Biol.}}
  \textbf{\bibinfo{volume}{85}}, \bibinfo{pages}{171--182}  (\bibinfo{year}{2010}).

\bibitem{Ref19}
\bibinfo{author}{Thompson, J. E.} et al.
\bibinfo{title}{Limits to catalysis by Ribonuclease A}.
\newblock \emph{\bibinfo{journal}{Bioorg. Chem.}}
  \textbf{\bibinfo{volume}{23}}, \bibinfo{pages}{471--481}
  (\bibinfo{year}{1995}).

\bibitem{Ref20}
\bibinfo{author}{Pross, A.} \& \bibinfo{author}{Khodorkovsky, V.}
\newblock \bibinfo{title}{Extending the concept of kinetic stability: Toward a
  paradigm for life}.
\newblock \emph{\bibinfo{journal}{J. Phys. Org. Chem.}}
  \textbf{\bibinfo{volume}{17}}, \bibinfo{pages}{312--316}
  (\bibinfo{year}{2004}).

\bibitem{Ref21}
\bibinfo{author}{Pascal, R.}, \bibinfo{author}{Pross, A.} \&
  \bibinfo{author}{Sutherland, J.~D.}
\newblock \bibinfo{title}{Towards an evolutionary theory of the origin of life
  based on kinetics and thermodynamics}.
\newblock \emph{\bibinfo{journal}{Open Biol.}} \textbf{\bibinfo{volume}{3}}, \bibinfo{pages}{130156}
  (\bibinfo{year}{2013}).

\bibitem{Ref22}
\bibinfo{author}{Kuramoto, Y.}
\newblock \bibinfo{title}{Self-entrainment of a population of coupled non-linear
  oscillators}. 
  \newblock In
  \emph{\bibinfo{booktitle}{International Symposium on Mathematical Problems in Theoretical Physics}} Vol. {\bibinfo{volume}{39}} (ed. \bibinfo{editor}{Araki, H.}) \bibinfo{pages}{420--422} (\bibinfo{publisher}{Springer Berlin Heidelberg}, \bibinfo{year}{1975}).

\bibitem{Ref23}
\bibinfo{author}{Hermoso~de Mendoza, I.}, \bibinfo{author}{Pach\'on, L.~A.},
  \bibinfo{author}{G\'omez-Garde\~nes, J.} \& \bibinfo{author}{Zueco, D.}
\newblock \bibinfo{title}{Synchronization in a semiclassical kuramoto model}.
\newblock \emph{\bibinfo{journal}{Phys. Rev. E}} \textbf{\bibinfo{volume}{90}},
  \bibinfo{pages}{052904} (\bibinfo{year}{2014}).

\bibitem{Ref24}
\bibinfo{author}{Makki, R.}, \bibinfo{author}{Mu\~nuzuri, A.~P.} \&
  \bibinfo{author}{Perez-Mercader, J.}
\newblock \bibinfo{title}{Periodic perturbation of chemical oscillators:
  Entrainment and induced synchronization}.
\newblock \emph{\bibinfo{journal}{Chem. Eur. J.}}
  \textbf{\bibinfo{volume}{20}}, \bibinfo{pages}{14213--14217} (\bibinfo{year}{2014}).

\bibitem{Ref25}
\bibinfo{author}{Simakov, D.~S.} \& \bibinfo{author}{P{\'e}rez-Mercader, J.}
\newblock \bibinfo{title}{Noise induced oscillations and coherence resonance in
  a generic model of the nonisothermal chemical oscillator}.
\newblock \emph{\bibinfo{journal}{Sci. Rep.}}
  \textbf{\bibinfo{volume}{3}}, \bibinfo{pages}{2404} (\bibinfo{year}{2013}).

\bibitem{Ref26}
\bibinfo{author}{Ramaswamy, R.} \& \bibinfo{author}{Sbalzarini, I.~F.}
\newblock \bibinfo{title}{Intrinsic noise alters the frequency spectrum of
  mesoscopic oscillatory chemical reaction systems}.
\newblock \emph{\bibinfo{journal}{Sci. Rep.}}
  \textbf{\bibinfo{volume}{1}}, \bibinfo{pages}{154} (\bibinfo{year}{2011}).

\bibitem{Ref27}
\bibinfo{author}{Bullara, D.}, \bibinfo{author}{De~Decker, Y.} \&
  \bibinfo{author}{Lefever, R.}
\newblock \bibinfo{title}{Nonequilibrium chemistry in confined environments: A
  lattice brusselator model}.
\newblock \emph{\bibinfo{journal}{Phys. Rev. E}} \textbf{\bibinfo{volume}{87}},
  \bibinfo{pages}{062923} (\bibinfo{year}{2013}).

\bibitem{Ref28}
\bibinfo{author}{Tompkins, N.} et al.
\newblock \bibinfo{title}{Testing Turing's theory of morphogenesis in chemical
  cells}. 
\newblock \emph{\bibinfo{journal}{Proc. Natl. Acad. Sci. USA}}  
  \textbf{\bibinfo{volume}{111}}, \bibinfo{pages}{4397--4402}
  (\bibinfo{year}{2014}).

\bibitem{Ref29}
\bibinfo{author}{De~Monte, S.}, \bibinfo{author}{d'Ovidio, F.},
  \bibinfo{author}{Dan{\o}, S.} \& \bibinfo{author}{S{\o}rensen, P.~G.}
\newblock \bibinfo{title}{Dynamical quorum sensing: Population density encoded
  in cellular dynamics}.
\newblock \emph{\bibinfo{journal}{Proc. Natl. Acad. Sci. USA}}
   \textbf{\bibinfo{volume}{104}},
  \bibinfo{pages}{18377--18381} (\bibinfo{year}{2007}).

\bibitem{Ref30}
\bibinfo{author}{Taylor, A.~F.}, \bibinfo{author}{Tinsley, M.~R.},
  \bibinfo{author}{Wang, F.}, \bibinfo{author}{Huang, Z.} \&
  \bibinfo{author}{Showalter, K.}
\newblock \bibinfo{title}{Dynamical quorum sensing and synchronization in large
  populations of chemical oscillators}.
\newblock \emph{\bibinfo{journal}{Science}}
   \textbf{\bibinfo{volume}{323}},
  \bibinfo{pages}{614--617} (\bibinfo{year}{2009}).

\bibitem{Ref31}
\bibinfo{author}{Schwab, D.J.}, \bibinfo{author}{Baetica, A.} \& \bibinfo{author}{Mehta, P.}
\newblock \bibinfo{title}{Dynamical quorum-sensing in oscillators coupled through an
  external medium}.
\newblock \emph{\bibinfo{journal}{Physica D}}
  \textbf{\bibinfo{volume}{241}}, \bibinfo{pages}{1782 -- 1788}
  (\bibinfo{year}{2012}).

\bibitem{Ref32}
\bibinfo{author}{Bar-Eli, K.}
\newblock \bibinfo{title}{On the stability of coupled chemical oscillators}.
\newblock \emph{\bibinfo{journal}{Physica D: Nonlinear Phenomena}}
  \textbf{\bibinfo{volume}{14}}, \bibinfo{pages}{242 -- 252}
  (\bibinfo{year}{1985}).

\bibitem{Ref33}
\bibinfo{author}{Ullner, E.}, \bibinfo{author}{Zaikin, A.},
  \bibinfo{author}{Volkov, E.~I.} \& \bibinfo{author}{Garc\'{i}a-Ojalvo, J.}
\newblock \bibinfo{title}{Multistability and clustering in a population of
  synthetic genetic oscillators via phase-repulsive cell-to-cell
  communication}.
\newblock \emph{\bibinfo{journal}{Phys. Rev. Lett.}}
  \textbf{\bibinfo{volume}{99}}, \bibinfo{pages}{148103}
  (\bibinfo{year}{2007}).

\bibitem{Ref34}
\bibinfo{author}{Lefever, R.}, \bibinfo{author}{Nicolis, G.} \&
  \bibinfo{author}{Borckmans, P.}
\newblock \bibinfo{title}{The Brusselator: It does oscillate all the same}.
\newblock \emph{\bibinfo{journal}{J. Chem. Soc. Faraday Trans. 1}}
  \textbf{\bibinfo{volume}{84}}, \bibinfo{pages}{1013--1023}
  (\bibinfo{year}{1988}).

\bibitem{Ref35}
\bibinfo{author}{Gray, P.}, \bibinfo{author}{Scott, S.~K.} \&
  \bibinfo{author}{Merkin, J.~H.}
\newblock \bibinfo{title}{The Brusselator model of oscillatory reactions.
  Relationships between two-variable and four-variable models with rigorous
  application of mass conservation and detailed balance}.
\newblock \emph{\bibinfo{journal}{J. Chem. Soc. Faraday Trans. 1}}
  \textbf{\bibinfo{volume}{84}}, \bibinfo{pages}{993--1011}
  (\bibinfo{year}{1988}).

   \bibitem{SR}
\bibinfo{author}{Benzi, R.}, \bibinfo{author}{Sutera, A.} \&  \bibinfo{author}{Vulpiani, A.}
\newblock \bibinfo{title}{The mechanism of stochastic resonance}.
\newblock \emph{\bibinfo{journal}{J. Phys. A: Math. Gen.}}
  \textbf{\bibinfo{volume}{14}}, \bibinfo{pages}{L453–L457}
  (\bibinfo{year}{1981}) 
  
   \bibitem{SR0}
\bibinfo{author}{McDonnell, M. D.} \&  \bibinfo{author}{Abbott, D.}
\newblock \bibinfo{title}{What Is Stochastic Resonance? Definitions, Misconceptions, Debates, and Its Relevance to Biology}.
\newblock \emph{\bibinfo{journal}{PLoS Comput. Biol.}}
  \textbf{\bibinfo{volume}{5}}, \bibinfo{pages}{e1000348}
  (\bibinfo{year}{2009}) 
  
  \bibitem{SR1}
\bibinfo{author}{Schneider, F. W.} \&  \bibinfo{author}{Munster, A. F.}
\newblock \bibinfo{title}{Chemical Oscillatlons, Chaos, and Fluctuations in Flow Reactors}.
\newblock \emph{\bibinfo{journal}{J. Phys. Chem.}}
  \textbf{\bibinfo{volume}{95}}, \bibinfo{pages}{2130--2138}
  (\bibinfo{year}{1991}).
  
  \bibitem{SR2}
\bibinfo{author}{Guderian, A.}, \bibinfo{author}{Dechert, G.},
  \bibinfo{author}{Zeyer, K.-P.} \&  \bibinfo{author}{Schneider, F. W.}
\newblock \bibinfo{title}{Stochastic Resonance in Chemistry. 1. The Belousov-Zhabotinsky Reaction}.
\newblock \emph{\bibinfo{journal}{J. Phys. Chem.}}
  \textbf{\bibinfo{volume}{100}}, \bibinfo{pages}{4437--4441}
  (\bibinfo{year}{1996}).

\bibitem{Ref36}
\bibinfo{author}{Zaikin, A.~N.} \& \bibinfo{author}{Zhabotinskii, A.~M.}
\newblock \bibinfo{title}{{Concentration wave propagation in two-dimensional
  liquid-phase self-oscillating system}}.
\newblock \emph{\bibinfo{journal}{Nature}} \textbf{\bibinfo{volume}{225}},
  \bibinfo{pages}{535--537} (\bibinfo{year}{1970}).

\bibitem{Ref37}
\bibinfo{author}{Elowitz, M.~B.} \& \bibinfo{author}{Leibler, S.}
\newblock \bibinfo{title}{A synthetic oscillatory network of transcriptional
  regulators}.
\newblock \emph{\bibinfo{journal}{Nature}}
\textbf{\bibinfo{volume}{403}},
 \bibinfo{pages}{335--338} (\bibinfo{year}{2000}).

\bibitem{Ref38}
\bibinfo{author}{Weitz, M.} et al.
\newblock \bibinfo{title}{Diversity in the dynamical behaviour of a
  compartmentalized programmable biochemical oscillator}.
\newblock \emph{\bibinfo{journal}{Nat. Chem.}}
  \textbf{\bibinfo{volume}{6}}, \bibinfo{pages}{295--302}
  (\bibinfo{year}{2014}).

\bibitem{Ref39}
\bibinfo{author}{Berkemeier, T.} et al. 
\newblock \bibinfo{title}{Kinetic regimes and limiting cases of gas uptake and heterogeneous reactions in atmospheric aerosols and clouds: A general classification scheme}.
\newblock \emph{\bibinfo{journal}{Atmos. Chem. Phys.}}
  \textbf{\bibinfo{volume}{13}}, \bibinfo{pages}{6663--6686}
  (\bibinfo{year}{2013}).

\bibitem{Ref40}
\bibinfo{author}{Nokkala, J.}, \bibinfo{author}{Galve, F.},
  \bibinfo{author}{Zambrini, R.}, \bibinfo{author}{Maniscalco, S.} \&
  \bibinfo{author}{Piilo, J.}
\newblock \bibinfo{title}{{Complex quantum networks as structured environments: Engineering and probing}}.
\newblock \emph{\bibinfo{journal}{{arXiv:1503.04635v1 [quant-ph]}}}  (\bibinfo{year}{{2015}}).

\bibitem{Ref41}
\bibinfo{author}{Zhang, W.-M.}, \bibinfo{author}{Lo, P.-Y.},
  \bibinfo{author}{Xiong, H.-N.}, \bibinfo{author}{Tu, M. W.-Y.} \&
  \bibinfo{author}{Nori, F.}
\newblock \bibinfo{title}{General non-Markovian dynamics of open quantum
  systems}.
\newblock \emph{\bibinfo{journal}{Phys. Rev. Lett.}}
  \textbf{\bibinfo{volume}{109}}, \bibinfo{pages}{170402}
  (\bibinfo{year}{2012}).
  
\bibitem{Ref42}
\bibinfo{author}{Pach\'on, L.~A.} \& \bibinfo{author}{Brumer, P.}
\newblock \bibinfo{title}{Direct experimental determination of spectral densities of molecular complexes}.
\newblock \emph{\bibinfo{journal}{J. Chem. Phys.}} \textbf{\bibinfo{volume}{141}},
  \bibinfo{pages}{174102} (\bibinfo{year}{2014}).

\end{thebibliography}
\end{document}